\documentclass[12pt]{article}
\usepackage[dvips]{epsfig}

\begin{document}
\setlength{\unitlength}{1mm}

\newcommand{\ba} {\begin{eqnarray}}
\newcommand{\ea} {\end{eqnarray}}
\newcommand{\be}{\begin{equation}}
\newcommand{\ee}{\end{equation}}
\newcommand{\n}[1]{\label{#1}}
\newcommand{\eq}[1]{Eq.(\ref{#1})}
\newcommand{\ind}[1]{\mbox{\tiny{#1}}}
\renewcommand\theequation{\thesection.\arabic{equation}}

\newcommand{\nn}{\nonumber \\ \nonumber \\}
\newcommand{\nl}{\\  \nonumber \\}
\newcommand{\pr}{\partial}
\renewcommand{\vec}[1]{\mbox{\boldmath$#1$}}

\author{Claude Barrab\`es\thanks{E-mail : barrabes@phys.univ-tours.fr}\\     
\small Laboratoire de Math\'ematiques et Physique Th\'eorique\\
\small  CNRS/UMR 6083, Universit\'e F. Rabelais, 37200 Tours, France\\
Valeri P. Frolov\thanks{E-mail : frolov@phys.ualberta.ca}\\
\small Theoretical Physics Institute, Department of Physics\\
\small  University of Alberta, AB, Canada, T6G 2J1\\
Emmanuel Lesigne\thanks{E-mail : emmanuel.lesigne@univ-tours.fr}\\
\small Laboratoire de Math\'ematiques et Physique Th\'eorique\\
\small  CNRS/UMR 6083, Universit\'e F. Rabelais, 37200 Tours, France }
\title{Geometric Inequalities and Trapped Surfaces in 
Higher Dimensional Spacetimes}
\date{}
\maketitle

\begin{abstract}
Geometric inequalities of classical differential geometry
are used to extend to higher dimensional spacetimes the
Penrose-Gibbons isoperimetric inequality and the hoop
conjecture of general relativity.

\end{abstract}
\thispagestyle{empty}

\section{Introduction}
\setcounter{equation}0

Higher dimensional spacetimes is now a common ingredient to most  
theories trying to unify gravity with the other forces of nature.
While in the earlier works the extra-dimensions were compact with a
size comparable with the Planckian scale, some recent  models
consider large \cite{ADD} or even infinite \cite{RS} extra
dimensions. In the so-called brane-world models the standard fields
are confined to a four-dimensional timelike hypersurface (the brane)
embedded in the higher dimensional spacetime (the bulk) where only
gravity can propagate. Black holes being the gravitational
solitons can be either attached to the brane or move in the bulk
space. The higher dimensional generalization of the vacuum black hole
solutions were obtained long time ago by Tangherlini \cite{Tang} for a
non-rotating black hole and by Myers and Perry \cite{Myer} for a
rotating one. With the
development of those models black holes in higher dimensional
spacetimes have come to play a fundamental role and  have received
much attention, see e.g. \cite{EmRe:02}--\cite{FrSt:03b} and
references therein. Another interesting
consequence of large extra-dimensions is the possibility of
production of mini black holes in high energy experiments such as
collision of ultrarelativistic particles in future colliders
or in cosmic rays, see e.g. \cite{EG,Land:02,Land:03} and
references therein.  The purpose of this paper is to use
geometric inequalities of classical differential geometry to derive
conditions for the formation of trapped surfaces and horizons in 
spacetimes with dimensions $D \geq 4$.

In ordinary general relativity where $D=4$ two inequalities
play an important role in 
the formation of horizons during a gravitational collapse. 
The first one is the so-called
Gibbons-Penrose ``isoperimetric'' inequality \cite{Pen}, 
\cite{Gib1} 
\be
{\cal A} \leq 4\pi\, (2GM)^2\, .\n{1}
\ee
It gives a relation between the area $\cal A $ of a trapped surface formed
during the collapse and the mass $M$ of the resulting black hole.
The second inequality arises in the formulation of the hoop conjecture
\cite{Tho} which states that  
{\it black hole with horizons form when and
only when a mass $M$ gets compacted into a region whose circumference
$\cal C $ in every direction is bounded by}
\be
{\cal C} \leq 2\pi\, (2GM)\, .\n{2}
\ee
The formulation of the hoop conjecture is rather vague as the type
of horizons is not specified and various interpretations can be
given to the mass and the cicumference.

In an attempt to produce counter-examples to the cosmic censorship
conjecture, Penrose \cite{Pen} 
considered a convex thin shell collapsing at the speed of light from infinity, 
in an initially flat spacetime, and he pointed that consistency with 
conventional theory demanded the validity of (\ref{1}). Gibbons \cite{Gib1}
noticed the connection between (\ref{1}) and the geometric  
Minkowski inequality \cite{Bur} 
\be
4\pi {\cal A} \leq Q^2\, ,\n{3}
\ee
which holds for any convex domain of $\mathbf{R}^3$,
$\cal A$ being the area of the boundary of this convex
and $Q$ its total mean curvature -see \cite{Gib2} for extension 
to the case of non-convex domains.
In the Penrose construction the mean curvature is shown to be
equal to $Q= 8\pi GM$ at the moment when the surface of the
collapsing shell gets trapped.
Once this value of $Q$ is introduced into (\ref{3})
the inequality (\ref{1}) immediately follows. In this approach
$M$ is the Bondi advanced mass (also equal to the ADM mass
and to the Hawking quasilocal mass) and is conserved during the collapse.   

The Penrose construction has also been used to give a more
precise formulation to the hoop conjecture. Calling $l$
the maximum circumference of plane projections
of the convex shell, and  $L$
the maximum circumference $L$ of plane curves drawn on the
surface of the shell, the following inequalities
\be
\pi L\, \leq\, 16\pi GM\,\leq 4l\, ,\n{4}
\ee
were derived in \cite{Bar1}.
Tod \cite{Tod1} found the slightly different result
\be
\pi l\, \leq\, 16\pi GM\,\leq 4l\, .\n{5}
\ee
which provides a stronger condition for the lower bound 
as $L\leq l$ for any convex.
According to say, (\ref{4}) the  necessary condition for
an apparent horizon to form is that the
mass $M$ gets compacted into a region such that $  L\, \leq\, 16\,GM\,$
Some examples satisfying this inequality were described in \cite{Bar2}, and
a numerical analysis of axisymmetric distributions by Chiba et al
\cite{Chib} led to the condition $C\,\leq\,15.8\,GM$
(their definition of $C$ does not exactly coincide with $L$).
Equality to the upper bound
in (\ref{4}) occurs for a sphere, and the lower bound is approached
when a cylinder ended by two hemispherical caps collapses 
toward its axis of symmetry to form a spindle singularity.

\section{Gibbons-Penrose inequality in higher dimensional spacetimes}
\setcounter{equation}0

Before extending the area inequality (\ref{1}) and the hoop
conjecture inequalities (\ref{4}) to higher dimensional spacetimes
let us briefly describe the Penrose-construction  in a $D$-dimensional
spacetime  - for a detailed derivation in the 4-dimensional case
see \cite{BI}. We consider a $(D-2)$-dimensional
convex thin shell which collapses at the speed of light from infinity
in an initially flat D-dimensional spacetime\footnote{It should be
emphasized that in the higher dimensional case there is no analogue
of the uniqueness theorem. In particular, one cannot exclude that the
topology of a higher dimensional black hole differs from the topology
of the sphere. A black ring solution \cite{EmRe:02} in 5-dimensional
case with the topology of the horizon $S^2\times S^1$ is one of the
examples. In the general case the existence of such solutions and
their stability is an open question \cite{EmMy:03}. 
In our approach we consider only
the black holes having spherical topology for the horizon.}.  
The history of the shell is a null hypersurface whose interior geometry
remains flat as long as the shell stays convex. As the shell implodes
outgoing light rays  emerging from the interior of the shell get more
and more focussed  when they cross the shell surface. A trapped
surface momentarily coincident with the shell forms when the
expansion rate of the outgoing light rays vanishes after crossing the
shell. Integrating the Raychaudhuri equation for the outgoing light
rays, accross the null hypersurface  at the moment of formation of
the trapped surface, one obtains the following relation 
\be 
K= 16\pi\, G_D\, \sigma\, ,\n{6} 
\ee 
between the the extrinsic curvature of the $(D-2)$-surface
of the shell (calculated in the interior Euclidean $(D-1)$-space),
and the   surface energy-density $\sigma$ of the shell. Here $G_D$ is
the gravitational constant of the $D$-dimensional spacetime. It is
worth mentioning that the main advantage of the Penrose construction
is that it allows to derive the relation (\ref{6})  without knowing
the spacetime geometry to the future of the imploding null shell. The total
mean curvature of the convex surface of the shell which is defined as
\be 
Q=\frac{1}{D-2} \int K\,dS\, ,\n{7} 
\ee 
becomes using (\ref{6})
\be 
Q=\frac{16\pi}{D-2}\, G_D\, M\, ,\n{8} 
\ee 
where $M$ is the Bondi
advanced mass of the shell which is conserved during the collapse and
is also equal to the $D$-dimensional analogues of the ADM mass and the
Hawking quasi-local mass. 

The generalization of the Gibbons-Penrose inequality (\ref{1})
to  $D$-dimensional spacetimes  
follows from the generalized Minkowski inequality
(see \cite{Bur} p.212) which states that for any closed 
convex $m$-dimensional surface immersed in $\mathbf{R}^n$, with
$2\leq m < n$, its area $A_m$ satisfies
\be
s_m \, (A_m)^{m-1} \leq Q^m\, .\n{9}
\ee
where $Q$ is the total mean curvature of the surface, and
$s_m$ is the area of the unit $m$-sphere
\be
s_n \,=\,\frac{2\,\pi^{\frac{n+1}{2}}}{\Gamma (\frac{n+1}{2})}\,\, ,\n{10}
\ee
$\Gamma (x)$ being the Euler function.
If one applies (\ref{9}) to
the imploding convex shell of the Penrose-construction,
then $m=D-2$, and using the expression (\ref{8}) for $Q$  
one obtains
\be
s_{D-2}\, (A_{D-2})^{D-3}\, \leq \, 
\left( \frac{16\pi G_D\, M}{D-2}\right)^{D-2}
\, .\n{11}
\ee
This inequality puts an upper bound on the area $A_{D-2}$
of any trapped
surface (this includes of course the apparent horizon) in terms
of the mass $M$ in a spacetime with $D \geq 3$ dimensions.
When $D=4$ one recovers the Gibbons-Penrose inequality
of general relativity,
$A_2 \leq 4\pi (2GM)^2 $, and for say, $D=5$ one gets
\be 
A_3 \leq \frac{32}{3} \left( \frac{2\pi}{3}\right)^{1/2} (G_5\, M)^{3/2}
\, .\n{12}
\ee
The dimension of the trapped surface increases with the dimension of
spacetime, recall that
$(A_{D-2})^{D-3}\, \sim\, (G_D\, M\,)^{D-2}\,
\sim\,(length\,)^{(D-2)(D-3)}$.
Inequalities involving the area of the boundary of plane
sections of the trapped surface with lower dimensions 
cannot be obtained from the method followed here.
One could introduce the maximum area $\Sigma_{max} $
of section of a convex of $\mathbf{R}^n$ with an hyperplane
by using the inequality $A_{n-1} \geq (b_n /b_{n-1})\,\Sigma_{max} $, where
$b_n $ is the volume of the unit $n$-ball (see \cite{Bur} p.152). 
However this
would not lower the exponents appearing in (\ref{11}) 
as $\Sigma_{max} $ has the same dimension as $A_{n-1}$, which
corresponds here to $A_{D-2}$ as $n=D-1$.

The Schwarzschild radius $r_H $ of a spherically symmetric
$D$-dimensional black hole with mass $M$ is equal to \cite{Myer}
\be
r_H \, =\,\left[ \frac{16\pi G_D M}{(D-2)s_{D-2}}\right]^{\frac{1}{D-3}}\, .
\n{13}
\ee
Using this relation 
the inequality (\ref{11}) takes the familiar form
\be
A_{D-2}\, \leq \, s_{D-2}\, r_H^{D-2}\, ,\n{14}
\ee
from which one immediately sees that equality occurs in the
spherical case. Another consequence of (\ref{11}) is that it can
be used to obtain an upper bound for the energy $E$ emitted as
gravitational radiation during the collapse of a mass $M$. 
If cosmic censorship holds and if a black hole is formed, then
the following quantity
\be
E_{max}\, =\, M\, -\,\frac{(D-2)\,(s_{D-2})^{\frac{1}{D-2}}}{16\pi\,G_D}\,
(A_{D-2})^{\frac{D-3}{D-2}}\, ,\n{15}
\ee
is always positive and yields the upper bound for $E$. 
In the case of brane-world models it is
known that the gravitational radiation will be 
emitted in the bulk as only gravitons can propagate in the bulk.

\section{Hoop inequalities in higher dimensional spacetimes}
\setcounter{equation}0

Let us now consider
the generalization to $D$-dimensional spacetimes of the 
inequalities (\ref{4}) associated with the hoop-conjecture. 
The following proposition will be used:\\
{\it Let $D$ be a convex domain of $\, \mathbf{R}^n$ and $Q$ 
the total mean curvature of the boundary $\partial D$ of $D$.
Let $\omega_{n-2} $ be the maximum area of the boundary of its
orthogonal hyperplane projections and $\Omega_{n-2} $ the maximum area
of $(n-2)$-dimensional sections of $\partial D$ by hyperplanes. 
Then the total mean
curvature satisfies the following inequalities}
\be
\frac{s_n}{2\,s_{n-2}}\, \Omega_{n-2}\, \leq \, Q\, \leq 
\, \frac{s_{n-1}}{s_{n-2}}\, \omega_{n-2}\, .\n{16}
\ee

{\em Proof:} Let $K$ be a convex domain embedded in $\, {\mathbf R}^n$, 
and $\partial K$ its boundary. Let $V_n (K)$ be the volume of $K$ 
and $A_{n-1}(\partial K)$ the area
$\partial K$. We define the $t$-expanded domain
$K_t$ associated with $K$ as $K_t = \{x \in {\mathbf R}^n\, |d(x,K)\leq t \}$
where $d$ is the Euclidean distance. The total
mean curvature $Q$ of $\partial K$ is such that
\be
Q\, =\,\left. \frac{1}{n-1}\, \frac{d\, A_{n-1}(\partial
K_t)}{dt}\right|_{t=0}\, .\n{18}
\ee
The Cauchy formula, \cite{Bur} p. 142, provides 
a relation between the area $A_{n-1}(\partial K)$ of the boundary 
$\partial K$,
and the mean volume $V_{n-1}(p_{\xi}(K))$ 
of the orthogonal plane projections of $K$ in arbitrary directions 
\be
A_{n-1}(\partial K)\, =\, \frac{1}{b_{n-1}}\,
\int_{\xi \in S_{n-1}}\, \,V_{n-1}(p_{\xi}(K))\, d\xi\, ,\n{19}
\ee
where $b_n $ is the volume of the unit $n$-ball, 
$\xi$ is a unit vector, $S_{n-1}$ is the unit-sphere and 
$p_{\xi}$ indicates the projection in the direction of $\xi$ 
onto a hyperplane orthogonal to $\xi$. The integral in (\ref{19}) 
is taken over the unit-sphere $S_{n-1}$.

One now applies the Cauchy formula to the $t$-expanded convex
domain $K_t$ and notices that $p_{\xi} (K_t)=(p_{\xi} (K))_t $. Then, 
using the following property
\be
\left.
\frac{d\, V_{n-1}(p_{\xi} (K_t))}{dt}\right|_{t=0}\, =\,
A_{n-2}(\partial p_{\xi}(K))\, ,\n{20}
\ee
where the r.h.s. of this equation represents the area of the
boundary of the projection of $K$, one gets
\be
Q\, =\, \frac{1}{s_{n-2}}\, \int_{\xi \in S_{n-1}}\, 
A_{n-2}(\partial p_{\xi}(K))\,d\xi \,  .\n{21}
\ee
The relation $s_{n-1}=n\, b_n $ between the
area  $s_{n-1}$ of the unit $(n-1)$-sphere and the volume $b_n$
of the unit $n$-ball has been used.

In order to derive the upper bound of (\ref{16}) one denotes   
$\omega_{n-2}\,\equiv \,\mbox{sup}_{\xi}\,
\lbrack A_{n-2}(\partial p_{\xi}(K))\rbrack$,
the maximum area of the boundary of the hyperplane
projections of $K$. Then the integration in (\ref{21}) yields
\be
Q\, \leq \,\frac{s_{n-1}}{s_{n-2}}\,\omega_{n-2}\, .\n{22}
\ee
For the lower bound of (\ref{16}) one considers the
intersection of $K$ with an arbitrary hyperplane $\Pi$. 
Let $\Sigma_{\Pi}$ be the closed plane  $(n-2)$-dimensional
domain resulting from the
intersection of $\Pi$ with the boundary $\partial K$ of $K$.
The following property \cite{Gra} applies to any
compact $q$-dimensional manifold $\cal V$ embedded in ${\mathbf R}^{p+q}$
with $p \geq 2$
\be
A_q ({\cal V})\,=\,\frac{s_p }{s_{p-1}\, s_{p+q}}
\int_{\xi \in S_{p+q-1}}\, \,A_q (p_{\xi}({\cal V}))\, d\xi\, .\n{23}
\ee
where, as above $p_{\xi}$ indicates the projection in the direction of 
the unit vector $\xi$ 
onto a hyperplane orthogonal to $\xi$. 
Applying this property to ${\cal V} =\Sigma_{\Pi}$ with 
$q=n-2$ and $p+q=n$, one gets
\be
A_{n-2} (\Sigma_{\Pi})\,=\,\frac{2}{s_n}
\int_{\xi \in S_{n-1}}\, \,A_{n-2} (p_{\xi} (\Sigma_{\Pi}))\, d\xi\, .\n{24}
\ee

As all the projections of $\Sigma_{\Pi}$ are boundaries of convex 
domains which are contained
within the projection of $\partial K$, one has 
$A_{n-2} (p_{\xi} (\Sigma_{\Pi}))\,\leq\, A_{n-2} (p_{\xi} (\partial K)$.
Introducing this into (\ref{24}) and using (\ref{21}) one gets
\be
\frac{s_n}{2}\,A_{n-2} (\Sigma_{\Pi})\,\leq\, s_{n-2}\, Q\, .\n{25}
\ee
This relation holds for any hyperplane $\Pi$ intersecting the convex 
domain $K$. Calling  
$\Omega_{n-2}\,\equiv\,sup_{\Pi}\,\lbrack A_{n-2}(\Sigma_{\Pi})\rbrack $, the
maximum area of plane $(n-2)$-dimensional sections of $\partial K$ 
by hyperplanes, one obtains
\be
\frac{s_n}{2\,s_{n-2}}\, \Omega_{n-2}\,\leq\,Q\, .\n{26}
\ee
The two results (\ref{22}) and (\ref{26}) 
provides the two geometric inequalities appearing (\ref{16})  
{\bf Q.E.D}.

The following relations derived from (\ref{10})
\be
\frac{s_n}{s_{n-2}}\,=\,\frac{2\pi}{n-1}\qquad ;\qquad
\frac{s_{n-1}}{s_{n-2}}\,=\,
{\sqrt {\pi}}\,\frac{\Gamma (\frac{n-1}{2})}{\Gamma (\frac{n}{2})}\, .\n{17}
\ee
could be used to present the proved inequalities in the form
\be
\frac{\pi}{n-1}\, \Omega_{n-2}\, \leq \, Q\, \leq 
\, {\sqrt {\pi}}\,\frac{\Gamma (\frac{n-1}{2})}{\Gamma (\frac{n}{2})}
\, \omega_{n-2}\, .\n{16a}
\ee

\subsection{Hoop of dimension $D-3$}

Let us apply  the inequalities (\ref{16}) to an imploding
null  shell of the Penrose construction  in a $D$-dimensional
spacetime. For this case
$n=D-1$ and $Q$ is given by 
(\ref{8}). It immediately follows that   
\be
\frac{s_{D-1}}{2s_{D-3}}\, \Omega_{D-3}\, \leq \, 
\frac{16\pi}{D-2}\, G_D\, M\, 
\leq \,\frac{s_{D-2}}{s_{D-3}}\, \omega_{D-3}\, ,\n{27}
\ee
where $\Omega_{D-3}$ and $\omega_{D-3}$ are defined above.
The left inequality in (\ref{27}) implies that a necessary
condition for the formation of a trapped surface is that a
convex body with mass $M$ gets compacted in a region such that  
the largest area $\Omega_{D-3}$ 
of a plane closed $D-3$-surface drawn on its boundary
satisfies, using (\ref{17})
\be
\Omega_{D-3}\, \leq \, 16\, G_D\, M\, ,\n{28}
\ee
which has the same form as the corresponding condition $L\,\leq\,16\,GM$ of
general relativity. A similar result was proposed by \cite{Ida}, see their
equation (53), and examples in 5-dimensional spacetimes were considered.

The right inequality in (\ref{27}) yields a sufficient condition
and it states that a trapped surface forms before the mass $M$ gets 
compacted into a region whose
orthogonal plane projections have a maximum area satisfying
\be
\omega_{D-3}\,\leq\,\frac{16\pi\,s_{D-3}}{(D-2)\,s_{D-2} }\, G_D\, M\, ,\n{29}
\ee
In terms of the Schwarzschild radius $r_H $ defined in (\ref{13}) 
the inequalities (\ref{27}) becomes 
\be
\frac{\pi}{D-2}\, \Omega_{D-3}\, \leq \, s_{D-3}\,r_H^{D-3}\, 
\leq \, \omega_{D-3}\, ,\n{30}
\ee
which shows, as in general relativity, that equality for 
the upper bound occurs in the case a sphere.

The inequalities (\ref{16}) and(\ref{27}) generalize to
spacetimes with $D$ dimensions
the relations (2) and (11) of \cite{Bar1} which were valid in
general relativity.
Let us mention that Tod's approach \cite{Tod1} 
does not seem to be straightforwardly generalizable    
to $\, \mathbf{R}^n$ with $n \geq 4$, i.e. to spacetimes
with dimension $D\geq 5$. 
As $G_D\, M \,\sim\,\Omega_{D-3}\,\sim\,\omega_{D-3} \,\sim\,(length)^{D-3}$,  
what was refered to as a hoop in general relativity becomes in fact 
a closed $(D-3)$-dimensional strip in $D$-dimensional spacetimes.
In the brane-world models only one of the $D-3$ dimensions of this strip 
belongs to the brane and the $D-4$ remaining ones 
correspond to the bulk.

\subsection{Hoop of dimension $D-4$} 
Introducing the circumference $\cal C$ 
of a curve and writing an inequality of the form
${\cal C} \leq  2\pi \, r_H$ giving a necessary and sufficient
condition to form horizons, as proposed in some
works \cite{Yos}, is not possible within our approach. 
One can however obtain for the upper bound of (\ref{16}) an expression
involving the area of a $(D-4)$-dimensional surface 
by repeating the projection procedure used in the derivation of (\ref{20}), 
and by using the isoperimetric inequality. 
Once the orthogonal projection $p_{\xi}$ of
convex domain $K$ is performed, the convex domain  $p_{\xi} (K)$ is all over
orthogonally projected in the direction of a new vector $\chi$
onto an hyperplane lying inside the hyperplane used in the $p_{xi}$ projection.
The domain $p_{\chi}(p_{\xi}(K))$ obtained after these projections 
is a $(n-2)$-dimensional domain which is
also convex. The Cauchy formula (\ref{19}) then gives 
\be
A_{n-2}(\partial p_{\xi}(K))\, =\, \frac{1}{b_{n-2}}\,
\int_{\chi \in S_{n-2}}\, \,V_{n-2}(p_{\chi}(p_{\xi}(K)))\, d\chi\, ,\n{31}
\ee 
and introducing this result into (\ref{21}) one obtains for the total mean
curvature
\be
Q\, =\, \frac{n-2}{s_{n-2}\,s_{n-3}}\, \int_{\xi \in S_{n-1}}\,
\int_{\chi \in S_{n-2}}\, 
V_{n-2}(p_{\chi} (p_{\xi}(K))) \,d\chi \,d\xi \,  .\n{32}
\ee
The next step is to apply to the domain $p_{\chi} (p_{\xi}(K))$ the 
isoperimetric inequality \cite{Bur} which states that between the
volume $V_n$ and the area $A_{n-1}$ of any convex of
${\mathbf{R}}^n $ one has
\be
n^n\, b_n\, V_n^{n-1}\,\leq \,A_{n-1}^n\, .\n{33}
\ee
One then gets a new upper bound for the total mean curvature $Q$
\be
Q\, \leq\, \frac{1}{s_{n-2}\,(s_{n-3})^{\frac{n-2}{n-3}}}\, 
\int_{\chi \in S_{n-2}}\,\int_{\xi \in S_{n-1}}\, 
A_{n-3}(\partial (p_{\chi} (p_{\xi}(K))))^{\frac{n-2}{n-3}} 
\,d\chi \,d\xi \,  .\n{34}
\ee
If one denotes by
$\sigma_{n-3}\,\equiv\,sup_{\chi ,\xi}\,
A_{n-3}(\partial (p_{\chi} (p_{\xi}(K))))$,
the maximum area of the boundary of two successive hyperplane 
projections of $K$, 
then the integration in (\ref{34}) yields
\be
Q\,\leq\, s_{n-1}\,\left( \frac{\sigma_{n-3}}{s_{n-3}}\right)
^{\frac{n-2}{n-3}}\, .\n{35}
\ee 
Using the expression (\ref{8}) for $Q$ and making $n=D-1$ one obtains
another expression for the upper bound of (\ref{16}) which now
involves the area of a $(D-4)$-dimensional surface. It is worth noticing that
this procedure only works for the upper bound 
and cannot be translated
to rewrite the lower bound of (\ref{16}) in a similar manner. 
Also it can be easily noticed that the procedure of
successive projections cannot be iterated more than twice as the
the Cauchy formula can no longer be used after two projections. 
In terms of the Schwarzschild radius $r_H$ the upper bound 
can be rewritten into the
simple form 
\be\label{A.15}
s_{D-4}\,(r_H )^{D-4}\,\leq \,\sigma_{D-4}\, ,\n{36}
\ee
which can be easily compared with the right inequality in (\ref{30}). 

As an example let us consider the case where $D=5$. In that case 
the relation (\ref{27}) gives
\be
\Omega_2\,\leq\,16\,G_5\, M\,\leq\, \frac{3}{2}\,\omega_2\, ,\n{37}
\ee
as $s_1 = 2\pi $, $s_2 = 4\pi $, $s_3 = 2\pi^2 $, and 
$s_4 = 8\pi^2 /3$. 
Suppose now that this situation applies to a 
brane-world model, and that the $2$-surface with area  $\Omega_2 $
has the form of an ellipsoid. 
Then one can write, omitting a factor of order unity, 
$\Omega_2 \simeq L\,l_5$, 
where $L$ corresponds to the 
size of the part of the ellipsoid lying on the brane
and $l_5$ in the bulk. The necessary condition to
form an apparent horizon is that the
product $ L\,l_5$ be small enough, i.e. the larger (smaller)
$l_5$ will be the smaller (larger) $L$ will have to be in order
to satisfy the inequality $\Omega_2\,\leq\,16\,G_5\, M\,$.
Also when $D=5$ two successive projections, as used in (\ref{35}), yields
a two dimensional domain whose perimeter has a maximum value $\sigma_1 $
which satisfies the inequality $2\,\pi\,r_H\,\leq\,\sigma_1 $.
This provides as mentionned earlier a sufficient condition to form 
a trapped surface. The condition takes a form which is apparently similar
to the inequality used in the hoop conjecture of general relativity,
except that we have now for the Schwarzschild radius,
$r_H^2\,=\,8\,G_5\,M\,/3\,\pi$ instead of $r_H\,=\,2\,G\,M$.

\noindent
\section*{Acknowledgment}\noindent
{The authors thank Professor Yuri Burago for helpful discussions.}
We are also very grateful to A. Gramain for providing the general
formula (\ref{23}).
The present work was partly supported by the NATO Collaborative
Linkage Grant (979723) and by the Centre National de la
Recherche Scientifique (France). One of the authors (V.F.) is
grateful to the Natural Sciences and Engineering Research Council of Canada
and to the Killam Trust for their financial support.

\end{document}